# NANOMETRIC SCALE SURFACE SCIENCE AND THE MARKOV CHAIN MONTE CARLO SIMULATION OF DISORDERED SYSTEMS

OLUWOLE E. OYEWANDE

Department of Physics, University of Ibadan, Ibadan, Nigeria.

Abstract

The Markov chain Monte Carlo method as a statistical mechanics technique for the study of macroscopic systems has furnished the scientific community with great knowledge and advances in the theory of phase transitions. While a number of Monte Carlo models have been proposed for the study of surface growth, these models have not nearly being studied as exhaustively as in the models of magnetic systems, a paradigm of which is the classical model of Ernest Ising. In particular, studies of phase transitions in surface/interface science at nanometric scales are almost non-existent. This article has been written to motivate research in this area of statistical mechanics from the perspective of surface science. In this article we survey the rudiments of the method along with some models of disordered systems such as magnetic systems, material fracture, nano-pattern formation under ion bombardment, and molecular chirality. We performed simulations of these models using the method and obtained results that are in excellent agreement with experimental observations.

## 1.     INTRODUCTION

The relevance and progress of theories of natural phenomena is strongly rooted in experiments that probe these phenomena under viable and accessible conditions. However, with increasing technological advances the experimental apparatus evolve with increasing sophistication often accompanied by huge operation or acquisition

---

e-mail: oe.oyewande@mail.ui.edu.ng

costs that create barriers which scientific enquiries must overcome before experiments designed to investigate them can be approved.

Fortunately, the increasing technological innovations have created a new avenue for experimental probes through the enablement of numerical simulations that mimic real-life experiments in the virtual world embedded in the central processing unit of a computer. Such simulations are based on the theoretical basis of the observed phenomena and, after verifying their results through comparison with experimental results, can be used to study areas not yet experimentally studied.

The Monte Carlo method is a probabilistic simulation method used in applying the theories of statistical physics to the study of macroscopic systems [1 – 3]. Such systems are termed "disordered" due to the large ($N \rightarrow \infty$) degree of freedom and the stochastic nature of their macroscopic events. Dynamics of surface nanostructure formation and evolution are commonly studied by means of stochastic partial differential equations. The sparse amount of simulation models are difficult to directly map to the continuum space and the statistical mechanics foundation of the solid-on-solid models are often left out in their formulation, most of which focus mainly on the calculation of the probabilities of the relevant events and the consequent simulation approaches without much recourse to the statistical mechanics background. This makes it difficult to effect the above mapping and leaves a number of unanswered questions.

This incompleteness in a number of reports, under the general assumption that they are either known or irrelevant, occurs in other areas as well. However, the statistical mechanics theory is more completely explored and developed in the application to magnetic systems and phase transitions [1, 2, 4, 5]. Some of these advances are as a result of what obtains from the application of the theory to seemingly unrelated areas (e.g self-organised criticality in granular mounds, biophysics, etc.), which have also benefited from the reverse feedback of applied results of the fostered advances in statistical theory.

# MARKOV CHAIN MC SIMULATION OF DISORDERED SYSTEMS

In this article we review the Markov chain Monte Carlo method in general as well as some models of disordered systems to which it has already been applied. The models considered are the Ising model of magnetic systems, a sputter-erosion model for studying pattern formation on material surfaces at nanometre length-scales, a model of material failure, and a model of molecular matching and packing. These models allow for the application of standard techniques of analysis in statistical physics. With recent advances in algorithm designs and combinatorial optimization in theoretical computer science [6, 7], the simulation algorithm for these models are being continually developed with increasingly faster and efficient simulations [4] which can help develop other fields and vice-versa in the ever thinning interface between the fields of scientific endeavors.

We start with a review of the basic but fundamental features of a computer simulation of disordered systems based on the theory of statistical physics. We start from standard material and gradually progress to research methods at the level of advanced texts and reference materials. In the context of this established theoretical simulation basis, we then discuss and present results of a few of the existing models referred to above, to facilitate the creation of new ones in yet unexplored areas. And, in particular, to motivate research towards the solution of open problems of statistical mechanics of disordered systems from the perspective of surface science and new frontiers in surface science (e.g. see Ref. [8])

## 2. REVIEW OF THE MARKOV CHAIN MONTE CARLO METHOD

The macroscopic or thermodynamic properties of a system can be calculated as a function of its microscopic properties through the relation (see e.g. Ref. [5, 9])

$$S = k_B \ln \Omega \qquad (1)$$

where the entropy $S$ is a macroscopic state variable, $\Omega$ is the number of microscopic states with energy in the range $E + dE$, and $k_B$ is the Boltzmann's constant. Note that the statistical definition of entropy given in (1) contains an implicit additive constant.

For macroscopic changes of state of the system under reversible (infinitesimal) processes the first law of thermodynamics states

$$d\mathfrak{E} = \sum_i \mathfrak{Y}_i \, d\mathfrak{X}_i \qquad (2)$$

where $\mathfrak{E}$ is the internal energy of the system, $\mathfrak{Y}_i$ are the generalized forces (e.g. intensive variables like pressure, surface tension, etc.), and $d\mathfrak{X}_i$ are the corresponding generalized displacements (e.g. extensive variables like volume, area, etc.) [5]. For irreversible processes the state variables $\mathfrak{E}$, $\mathfrak{Y}_i$, and $\mathfrak{X}_i$ are undefined. For processes in which the relevant state variables are, for instance, $\mathfrak{S}$, the volume $\mathfrak{V}$, and the number of moles $\mathfrak{N}$, then $\mathfrak{E} = \mathfrak{E}(\mathfrak{S}, \mathfrak{V}, \mathfrak{N})$ is the function of state and according to (2) the infinitesimal change in the entropy is given by

$$d\mathfrak{S} = \frac{1}{\mathfrak{t}} d\mathfrak{E} + \frac{\mathfrak{p}}{\mathfrak{t}} d\mathfrak{V} - \frac{\mu}{\mathfrak{t}} d\mathfrak{N} \qquad (3)$$

to which an application of Euler's relation gives

$$\mathfrak{S} = \frac{\mathfrak{E}}{\mathfrak{t}} + \frac{\mathfrak{p}\mathfrak{V}}{\mathfrak{t}} - \frac{\mu\mathfrak{N}}{\mathfrak{t}} \qquad (4)$$

Depending on the nature of the interaction of the system with its environment there may be one or more of energy exchange, matter exchange, or exchange of spatial extent, which determines the statistical ensemble, i.e. assembly of microstates subject to one or more of these constraints, within which the system evolves and is simulated. In any such scenario the equilibrium state of the system is any of a large number of equally probable microstates, whereas the probability of a microstate varies between equilibrium states. In fact the number of microstates is infinite if the phase space points form a continuum.

# MARKOV CHAIN MC SIMULATION OF DISORDERED SYSTEMS

For reversible processes carried out on a canonical ensemble where $\mathfrak{V}$ and the number of particles $N$ $(N \propto n)$ are constant, $d\mathfrak{V} = dN = 0$ and the number of microstates $\Omega_i$ with energy $\mathcal{E}_i$ at temperature $\mathfrak{t}$ is given by

$$\Omega_i = \exp\left(\frac{\mathcal{E}_i}{k_B \mathfrak{t}}\right) \qquad (5)$$

The relationship between $\mathfrak{E}$ and $\mathcal{E}_i$ is that between the average of a data set and the elements of the data set; as we shall see below the system must be allowed to equilibrate at a particular temperature before measurements are made, nevertheless, the energy fluctuates about an average value after the equilibration time such that $\mathfrak{E} = \omega^{-1} \sum_j \mathcal{E}_j$; where $\omega$ is the number of microstates with (possibly different) energies $\mathcal{E}_j$ through which the system evolved in the measurement process.

According to the fundamental postulate of statistical all the microstates corresponding to a particular macroscopic state are equally probable, thus the probability $P_i$ of a microstate with energy $\mathcal{E}_i$, $P_i \propto 1/\Omega_i$, i.e.

$$P_i \propto \exp\left(\frac{\mathcal{E}_i}{k_B \mathfrak{t}}\right) \qquad (6)$$

or

$$P_i = \frac{1}{\mathcal{Z}} \exp\left(\frac{\mathcal{E}_i}{k_B \mathfrak{t}}\right) \qquad (7)$$

where the constant $\mathcal{Z}$ that ensures the normalization of the probabilities $P_i$, so that $\sum_i P_i = 1$, is the so-called partition function given by

$$\mathcal{Z} = \sum_i \exp\left(-\frac{\mathcal{E}_i}{k_B \mathfrak{t}}\right) \qquad (8)$$

Note that $\Omega =$ constant for the micro-canonical ensemble, which implies $P_i =$ constant for this ensemble, hence, such states can be sampled uniformly as in Monte Carlo integration and not by importance sampling as is the case in the canonical ensemble.

The requirement that the macroscopic states be reversible imposes the condition of ergodicity on the microscopic phase space of this system such that every microstate must be accessible along a finite path or trajectory that connects the microstates in phase space. This is achieved by constraining the simulated microstates to a Markov chain such that a new state is attained from a previous one in the chain. Thus in simulations of disordered systems in a given statistical ensemble the goal is to visit the phase points or ensemble microstates $m$ with a probability $P_i$ proportional to some given distribution $\rho(m)$, where the phase points are locally correlated in a sequence of points $m_1, m_2, m_3, \cdots$ each derived from the subsequent one and all constituting a Markov chain.

In the canonical ensemble $\rho(m_i)$ is given by $\exp\left(-\frac{\varepsilon_i}{k_B T}\right)$ and $P_i$ is given by (7). $\rho(m)$ is not necessarily normalized to have unity integral or sum over the sampled region but is proportional to a probability. Now $P_i$ is a conditional probability, i.e. the probability of accessing the microstate $m_i$ given that the system is in the microstate $m_{i-1}$

$$P_i = P(m_i | m_{i-1}) \propto \rho(m_i) \qquad (9)$$

So $P_i$ are transition probabilities of transition from microstate $m_{i-1}$ to $m_i$. For us to obtain the partition function we would have to first sample all microstates and evaluate the sum of the distribution functions $\rho(m_i)$ which is, however, unnecessary since we must have

$$\frac{P(m_i | m_{i-1})}{P(m_{i-1} | m_i)} = \frac{\rho(m_i)/Z}{\rho(m_{i-1})/Z} \qquad (10)$$

that is,

$$\rho(m_{i-1}) P(m_i | m_{i-1}) = \rho(m_i) P(m_{i-1} | m_i) \qquad (11)$$

The condition (11) is known as detailed balance. It allows us to calculate transition probabilities in terms of relative transition probabilities between successive microstates without actually calculating the partition function. For instance, in the canonical ensemble (7) and (10) imply

$$P_{m_{i-1} \to m_i} = \exp\left(-\frac{\Delta\mathcal{E}}{k_B t}\right) \qquad (12)$$

where $\Delta\mathcal{E} = \mathcal{E}_i - \mathcal{E}_{i-1}$ is the energy difference in a transition from the microstate $m_{i-1}$ with energy $\mathcal{E}_{i-1}$ to the next microstate $m_i$ with energy $\mathcal{E}_i$ in the Markov chain, and $P_{m_{i-1} \to m_i} = \frac{p(m_i|m_{i-1})}{p(m_{i-1}|m_i)}$ is the relative conditional probability that the system will evolve to microstate $m_i$ from the microstate $m_{i-1}$.

According to (12) the system will certainly change into the microstate $m_i$ from $m_{i-1}$ if $\mathcal{E}_i \leq \mathcal{E}_{i-1}$ since in this case $P_{m_{i-1} \to m_i} \geq 1$, whereas it will likely remain in the microstate $m_{i-1}$ if $\mathcal{E}_i > \mathcal{E}_{i-1}$ since $P_{m_{i-1} \to m_i} < 1$ in this case and $P_{m_{i-1} \to m_i} \to 0$ as $\Delta\mathcal{E} \to \infty$. This ensures that the vast phase space is "importance" sampled in such a way that the system evolves in the few most probable microstates in the phase space in the course of the simulation and reduces the rigor and inefficiency of having to go over or uniformly sample ("simple" sampling) the entire phase space in the calculation of average quantities such as, e.g.

$$\langle\mathcal{E}\rangle = \frac{\sum_i \mathcal{E}_i \exp(-\beta\mathcal{E}_i)}{\sum_i \exp(-\beta\mathcal{E}_i)} \qquad (13)$$

In the calculation of (13) most of the Boltzmann weights $\exp(-\beta\mathcal{E}_i)$ are vanishingly small, corresponding to the very large number of microstates in which the system is unlikely to be found, and therefore contribute very little to the average value. Note that, depending on the application, $\mathcal{E}$ and $T$ may have entirely different physical meaning.

## 2.1 Simulation Algorithm

Based on this theory, a Markov chain MC simulation of the evolution of the microstates of a system in statistical mechanics can be performed by using the following algorithm, known as the Metropolis algorithm: Generate a trial configuration $m_{i-1}$

$$m_{i-1} = \{\sigma_1, \sigma_2, \cdots, \sigma_N\} \qquad (14)$$

for the initial microstate of the $N$-particle system and obtain the next microstate $m_i$ in the Markov chain by tweaking the coordinate $\sigma_k$ of a single ($k$th) particle. Calculate the acceptance probability $\alpha(m_{i-1}, m_i)$ of the new microstate $m_i$ using the formula

$$\alpha(m_{i-1}, m_i) = \min(1, P_{m_{i-1} \to m_i}) \qquad (15)$$

This means that the transition $m_{i-1} \to m_i$ is accepted if $\mathcal{E}_i \leq \mathcal{E}_{i-1}$ $[\alpha(m_{i-1}, m_i) = 1]$ otherwise it is accepted with a probability $[\alpha(m_{i-1}, m_i) = P_{m_{i-1} \to m_i} < 1]$ which, at constant $T$, decreases with increasing $\Delta\mathcal{E}$.

This algorithm therefore ensures that the system always tend towards microstates of lower energy than the previous one, but unlike the so-called "greedy algorithms" [10] that never accept a higher energy option, it also allows the system to sometimes (though with very low probability) accept a move to a higher energy microstate, wherein lies its power and vast applications to diverse optimization (minimization/maximization) problems. For instance, if higher energy moves are always rejected then there will be no possibility for the system to move out of any of the local minima in Fig. 1 and the system would therefore not equilibrate to the global minimum. However, as the temperature decreases the probability of the system moving out of a local minimum, if trapped there at low $T$, also decrease and at such low $T$ it may be difficult to attain a microstate of global energy minimum; depending on the energy landscape of the system.

It is best to run independent simulations with different initial conditions, e.g. one with a cold start ($T = 0$) and another with a hot start ($T = \infty$), so as to avoid inaccurate results due to local minima pitfalls. For the cold start a starting configuration or microstate that is known to have the lowest energy is used, while for the hot start a configuration that corresponds to the highest energy is used. The result of the independent simulations must be the same irrespective of the starting configuration, otherwise one or both must have ended in local minimum and further checks are necessary.

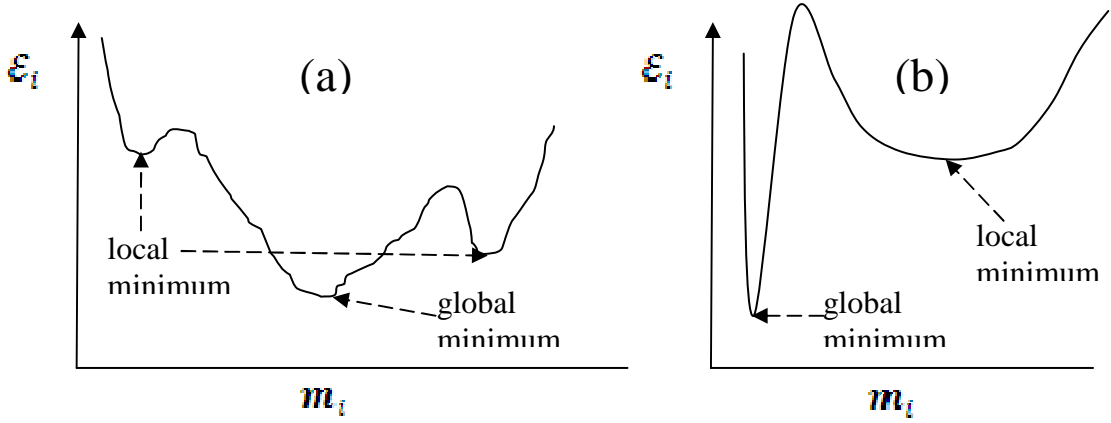

Figure 1: Hypothetical energy landscape of two systems. In (b) the global minimum is within a very narrow region of the phase space of the system which means that reaching it may be almost impossible if trapped elsewhere at low $T$.

Since the system equilibrates to a "global" energy minimum in which it then spends the rest of its time, the probability $P(m_i|m_{i-1})$ is time dependent , i.e. $P(m_i|m_{i-1}) = P_{m_i}(t)$ and its time dependence is governed by the master equation

$$\frac{\partial P_{m_i}(t)}{\partial t} = \sum_{m_j \neq m_i} \left[ P_{m_j}(t) R_{m_j \to m_i} - P_{m_i}(t) R_{m_i \to m_j} \right] \qquad (16)$$

where $P_{m_i}(t)$ is the probability of the system being in the microstate $m_i$ at time $t$, and $R_{m_i \to m_j}$ is the rate of transition from microstate $m_i$ to $m_j$. In equilibrium $\frac{\partial P_{m_i}(t)}{\partial t} = 0$ and $R_{m_j \to m_i} = R_{m_i \to m_j}$.

Obviously the efficiency of this algorithm, at the so-called "importance sampling" of the phase space region in which the system spends most of its time, will depend on the energy landscape of the system; the algorithm will be more efficient and faster for a system with the energy landscape of Fig. 1(a) than that of Fig. 1(b). As a result, a number of cluster algorithms have been proposed to increase the simulation speed and efficiency at "importance sampling" by avoiding the slow changes of state that occur with time in certain stages of the simulation (e.g. critical slowing down, frustration). Such algorithms consider microstate transitions involving clusters of particles instead of single particles [11 – 14].

Since the system may end up in local minima it is better to let the simulation run for a while before taking measurements so as to allow the system to equilibrate; the time required for the system to equilibrate is known as the equilibration time, which depends on the energy landscape of the system. The configurations, or microstates, are usually stored as arrays whose elements are the single-particle states corresponding to each particle of the system.

The energy of a microstate depends on the interactions between the particles and is the sum of the energies of these interactions. The particles, being indistinguishable (e.g. fermions or bosons), must have identical surroundings and interactions in such a way that the boundary particles are not distinct from the inner ones. This is achieved by imposing periodic boundary conditions on the system, thus transforming the $D$-dimensional system into a $(D + 1)$-dimensional torus, so that the particles at one end of the system sees the particles at the other end as nearest neighbors. The storage requirements for specific problems or the geometrical nature of certain systems may require other boundary conditions like screw-periodic, anti-periodic, mean-field, or free-edge etc., boundary conditions. Other lattice structures may also be more appropriate for specific problems [1, 2].

## 2.2 Uniform Deviates

The recipe for simulation is incomplete without (pseudo) random numbers [10]. We can not choose $\sigma_k$ in a deterministic version because that would be unrealistic, it would either imply a prior knowledge of the phase trajectory followed by nature or that we are the ones dictating what course nature would take which is impossible for physics in the subatomic realm. Hence, $\sigma_k$ is chosen randomly by using floating-point uniform deviates such that each $\sigma_k$ has the same likelihood of being picked as any other. Our knowledge of the probability of a trial microstate is also not enough because we need to know when to accept or reject it according to its probability. Again, a floating-point uniform deviate is used.

For instance, if the probability of an event is 0.1 then the event will occur once out of every 10 attempts, as the number of attempts tends to infinity. To simulate this uniform deviates are used. The random numbers are assumed to be uniformly spread over an interval (0 to 1 for floating-point random numbers) such that each random number has an equal probability of being chosen as any other one.

## 2.3 Finite Size Effects

The number of particles (system sizes) that may be considered in a simulation is much less than the number of particles in a typical macroscopic system, i.e. the simulated system size $N \ll \infty$ whereas in the thermodynamic limit we should have $N \to \infty$. Typically, real $N \sim 10^{23}$ particles whereas simulation $N \sim 10^2$. For instance, a system with only two possible single-particle states will have $2^N \sim 2^{10^{23}}$ microstates or configurations while the simulation phase space comprises of $\sim 2^{100}$ microstates.

However, use of small $N$ still gives very accurate results when compared with real systems for which $N \to \infty$ provided the simulation is for temperatures far from the critical temperature at which a phase transition occurs. For temperatures close to and above the critical temperature $T_c$ the effects of using small $N$, known as finite size effects, become very strong and the simulation results become inaccurate; the inaccuracy increases with decreasing $N$. Hence, for studies involving phase transitions, or for simulations at temperatures above $T_c$, it is important to do a finite size scaling of the results to obtain the actual results of a real thermodynamic system.

The form of the finite size scaling will depend on the set of thermodynamic quantities being measured and their scaling relations (1) (2) (3). It also depends on the order parameter, which is a quantity that has a value of zero on one side of the phase transition and a nonzero value on the other side of the phase transition; i.e. which is like a step function whose step occurs at the phase transition.

The origin of the finite size effects is in the divergence of the correlation length $\xi$ (the length scale of the fluctuations in the order parameter) at the critical temperature as a result of which it is impossible to get accurate results of the thermodynamic properties of the system at $T_c$ from any simulation performed on a finite lattice. Since the lattice size is less than the value of $\xi$ at $T_c$ all the lattice points are correlated and one is unable to obtain the scenario of linearly independent single-particle states.

## 3.0 MODELS OF SOME DISORDERED SYSTEMS AND BASIC SIMULATION RESULTS.

In this section, we present some general simulation models which are typical examples of the successes of the Markov chain Monte Carlo approach to the simulation of the complex, disordered, systems frequently encountered in nature which are of great technological importance.

### 3.1 Magnetism and the Ising Model

Magnetism is one of the most studied phenomena in materials. It has been a study of active and intense research interest for centuries [15, 16]. Any material is composed of atoms bound together by cohesive van-der-Waals, covalent, electrovalent, or metallic (electron gas) interactions which reduce to electrostatic interactions between dipoles made up of the negatively charged electrons on one end and the positively charged ion cores on the other. Each electron has an intrinsic, purely quantum mechanical, angular momentum called the spin $S$ of the electron. Associated with each electron spin is a magnetic moment of magnitude $\frac{e}{mc}S$ (i.e. noting that the gyro-magnetic ratio $\frac{e}{mc}$ in this case is twice the ratio of the magnetic moment to the angular momentum) which is solely responsible for the magnetism of the material; $e$ is the electronic charge, $m$ is the mass of the electron, and $c$ the speed of light in vacuum. Thus the magnetic property is a quantum mechanical property and does not have a classical analog, which means that the Hamiltonian $\mathfrak{H}$ of the system does not include position and linear momentum operators.

The exchange energy due to spin-spin interactions between spin $i$ and spin $j$ can be specified as $\frac{e^2}{m^2c^2r_{ij}}S_iS_j$ where $r_{ij}$ is the distance between the spins. For a magnetic material in a magnetic field, therefore, the only changes in the energy of the system are those due to the interaction of the spins with each other and with the external magnetic field $H$. Hence, the Hamiltonian $\mathfrak{H}$ of the system is given by

$$\mathfrak{H} = \frac{e^2}{m^2c^2}\sum_{j>i}\sum_i \frac{1}{r_{ij}}S_iS_j - \frac{e}{mc}\sum_i HS_i \qquad (17)$$

This is equivalent to the model of Ernest Ising for which, for simulations on lattices of unit spacing with nearest neighbor interactions, the eigenvalues of the Hamiltonian are expressed as

$$E = -J\sum_{\langle i,j \rangle}S_iS_j - H\sum_i S_i \qquad (18)$$

where $J$ is the unit of the spin-spin exchange interaction and is often chosen to be $+1$, in which case the lowest energy state will tend to have neighbouring spins aligned and the low temperature ground state will be a ferromagnet (all spins aligned), or $-1$ in which case the lowest energy state will tend to have neighboring spins opposed in orientation and the low temperature ground state will be an anti-ferromagnet (alternating spins). $\langle i, j \rangle$ denotes a nearest neighbor pair, and $H$ is the unit of the spin-field coupling.

The spin operators (actually $S = S_z$ for each spin, assuming the field is oriented along the $z$-axis so that $H = H_z$) have eigenvalues $\pm \frac{1}{2}\hbar$ where $\hbar = \frac{h}{2\pi}$ is the Dirac action constant. In simulations the eigenvalues of the spin operator $S$ are chosen to be $\pm 1$. Thus the number of single-particle microstates of the system is reduced to only two; for up and down orientations of the intrinsic spin.

When the exchange interaction energy $J$ varies from bond to bond then we have a spin-glass system which is widely studied with a variety of interesting ramifications [17].

## 3.2 Surface Nanostructure Formation: Results of Simulation with the HKGK Model.

A more recent observed phenomenon in surface science, than the magnetism of condensed matter described above, is the formation of self-organized nano-patterns on material surfaces when bombarded by a beam of energetic particles, with energies in the range of keV [8, 18 – 23]. The surface morphology arises as a result of the interplay between the roughening processes of stochastic removal of surface particles by ion-sputtering, and the smoothening processes of diffusion of surface particles or radiation-induced viscous flow. The roughening process creates instability of the surface against further perturbations by the eroding ions such that the sputter yield becomes curvature dependent with the result that surface depressions are eroded in preference to surface protrusions. According to the continuum theory, it is this curvature dependence that leads to the formation of ripples of wavelength $\lambda = 2\pi\sqrt{2K/|\nu|}$, where $K$ is the surface diffusivity, and $\nu$ is a surface tension coefficient. The wavelength of the ripples is on the nanometer scale and they are oriented along the direction with the largest $|v|$ [21].

The experimentally observed ripple orientation is parallel to the projection of the ion beam direction onto the surface plane for large incidence angles (close to grazing incidence), and perpendicular to the projection for small incidence angles; except for metallic surface with anisotropic diffusion in which case the ripple orientation is perpendicular to a crystallographic direction (i.e. the one favored for diffusion) at small incidence angle.

Hartmann, Kree, Geyer, and Koelbel (HKGK) proposed a discrete solid-on-solid Monte Carlo model [24] for the simulation of surface modification and evolution by ion-bombardment which, in contrast to the Ising model above, simulates the surface morphology indirectly by resorting to the effect of collision cascades on the surface evolution without the direct need for a surface Hamiltonian. These collision cascades are set off in the near sub-surface layer by the impinging ion and are responsible for the redistribution of the energy of the impinging ion to the surface particles. The energy $E(x)$ received by a surface particle at position $x = (x_1, x_2, x_3)$ due to the arrival of the ion somewhere in the vicinity of this surface particle is assumed to be of the Gaussian form [25]:

$$E(x) = \frac{E}{\left(\sqrt{2\pi}\right)^3 \alpha \rho^2} \exp\left(-\frac{x_3^2}{2\alpha^2} - \frac{x_1^2 + x_2^2}{2\rho^2}\right), \qquad (19)$$

where $\alpha$ and $\rho$ are the widths of the collision cascade ellipsoid parallel and perpendicular to the ion beam direction, respectively. A sputtering process is simulated by eroding surface particles in the vicinity of the impact position of the ion with probabilities proportional to $E(x)$ [26 – 28], and a diffusion or smoothing process is simulated by using any of the available diffusion models (e.g. [29, 30]), depending on the experimental hopping rates for the material concerned.

We performed extensive simulations using this model and the surface profiles obtained from a typical simulation are shown in Fig. 2. As can be seen from this figure, the simulation model accurately generates the observed surface ripples and their orientation with respect to the ion beam direction and ion incidence. The projection of the ion beam direction onto the surface is represented by the short thin bar on the profiles. These results are in excellent agreement with the experimental observations [26 – 28] of the ripple orientation being perpendicular to the ion beam direction for incidence angles less than a threshold angle determined to be about 70°. They are also in excellent agreement with the experimental results that the ripple amplitude grows with sputter time; this is shown by the increasing magnitude of the surface height range on the vertical grey scale of the figure.

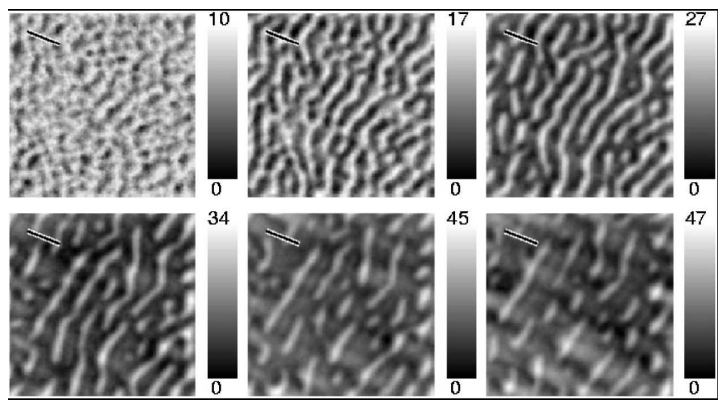

Figure 2: Surface profiles for different sputtering times (Monte Carlo steps). From top to bottom, left to right: t = 0.5, 1.5, 4, 9, 14, and 20 ions/atom. The vertical scale indicates the surface height range. The ion beam direction is indicated by the bar on the profiles, showing that the ripple orientation is perpendicular to ion beam direction as expected since $\theta = 50°$ here.

### 3.3 Molecular Chirality and the YNL Matching Algorithm: Simulation Results.

The problem of measuring the extent to which an object fails to coincide with or match its mirror image [31 – 39] when optimally superimposed on the mirror image was recently tackled in a direct way through simulations of the matching process using the Markov chain MC method [40] (YNL algorithm). The measure of chirality or degree of matching of the objects (an object and its image or two different objects as in molecular similarity or facial resemblance) that have been optimally superimposed by the YNL algorithm is based on the concept of the Hausdorff distance between sets as proposed by Buda and Mislow [33].

This direct approach entails the simulation of the matching process which includes a large number of rotations and translations such that the Hausdorff distance $d_H$ between the objects being matched is as small as possible. $d_H = 0$ if an object is achiral or if two objects match, and $d_H > 0$ if an object is chiral or two objects do not match. The challenge is that one may get very very close to the global minimum of $d_H$ only to be pushed into a completely different region of the $d_H$ landscape with the slightest of translations in the matching process. But the YNL Monte Carlo matching algorithm [40] always arrive at the optimal match. Hence, this algorithm can be of immense benefit to the handling of harmful enantiomorphs of useful chiral molecules in drug design, to database searches in the area of sequence alignment and queries on other databases, and to the understanding of the mechanism of chirality transfer from a chiral dopant to an achiral host.

The YNL algorithm searches for the optimal superimposition of an object $A$ and its mirror image $A' = -A$, or second object, by performing a series of random rotations and translations of $A$ keeping $A'$ stationary. The simulation is started with a random orientation and translation of $A$ which represents the initial state or configuration. The matching process is iterated down over the 'temperature', thus annealing the process and ensuring that as the system cools, the rotation and translation reduces in magnitude so that at high 'temperature' the system moves freely from one region of the $d_H$ landscape to another in the search for the region that contains the global minimum, and much less freely when this region is supposed to have been found (at low 'temperature'). During each iteration stage over the temperature, it is kept fixed while $N_{MC}$ Monte Carlo steps of the matching process are performed.

In each MC step the state of the system, which is the orientation and position of $A$ relative to $A'$, is changed by rotating and translating $A$ and the move is accepted according to the standard MC transition acceptance rules, tailored towards acquiring minimum $d_H$ or $d_{H_{min}}$. $d_{H_{min}}$ provides a measure of the chirality $\chi_H$ of the object through the chirality measure of Buda et al. [33]

$$\chi_H(A) = \frac{d_{H_{min}}(A, A')}{d(A)} \qquad (20)$$

where $d_{H_{min}}(A, A')$ is the minimum Hausdorff distance between $A$ and $A'$ over all positions and orientations of $A$ and $A'$, $d(A)$ is the diameter of $A$. The results of our simulations as applied to the multidimensional search for the most chiral tetrahedral plane shapes with $D_2$, $C_2$, and $C_1$ symmetries are presented in Table 1. As can be seen from the table, our results are in excellent agreement with the results of Buda, Auf der Heyde, and Mislow [32, 33] who obtained their results by using another, albeit more tedious, numerical optimization method.

**Table 1: A summary of our (YNL) results of the most chiral tetrahedral, and their comparison to the results of Mislow and co-workers (BHM).**

| Symmetry | Method | Internal angles | $\chi_H$ | Relative to BHM |
|---|---|---|---|---|
| $D_2$ | BHM | (35.1, 60.5, 35.1, 84.4) | 0.221 | 1 |
| | YNL | (35.01, 60.51, 35.01, 84.44) | 0.221 | 1 |
| $C_2$ | BHM | (45.6, 58.5, 38.0, 34.7) | 0.252 | 1 |
| | YNL | (45.23, 58.62, 37.92, 35.24) | 0.253 | 1.004 |
| $C_1$ | BHM | (44.4, 59.7, 37.7, 36.0) | 0.248 | 1 |
| | YNL | (45.15, 58.55, 37.74, 35.12) | 0.253 | 1.02 |

### 3.4 Material Fracture and the Fiber Bundle Model

Fracture in heterogeneous materials is a complex physical problem which has been a focus of research interest for a long time [41 – 43]. Fiber bundle models (FBMs) form a fundamental class of approaches to the fracture problem through their capture of the essential features of material breakdown. They are models of materials as composites of fibers such that the breakdown of the material is as a result of the propagation of fracture among the failing fibers. One such model which provides a deep understanding of the intrinsic nature of the fracture process is the dynamic fiber bundle model [44]. According to the dynamic FBM the material breaks down as a result of the fatigue of its fibers over time. That is, when the material is stressed or loaded the fibers share the load and get progressively stressed with time such that even if the external load or source of stress is removed the acquired fatigue in the fibers remain and are incremented with the introduction of a new source of stress until the individual fibers reach their threshold and fail, leading to the overall failure of the material with time. This differs from the static FBM [45] which attributes the failure of the individual fibers to their quasi-static loading due to the gradual increase in the external load.

The dynamic FBM includes a normalised combined load sharing rule [44]

$$S(\gamma, r_{ij}) = \frac{1}{r_{ij}^{\gamma}} \frac{1}{\sum_{i \in F} r_{ij}^{-\gamma}} \qquad (21)$$

$r_{ij}$ is the distance between an active fiber $i$ and a failed fiber $j$, $F$ denotes the set of active fibers, and $\gamma$ is a variable parameter that controls the effective range of interaction among the fibers. This combines the local load sharing rule ($\gamma = \infty$) in which the load borne by a failed fiber is shared equally among the four nearest neighbour fibers, the global load sharing rule ($\gamma = 0$) in which the load of a failed fiber is shared among all existing fibers, and the variable range load sharing rule ($0 < \gamma < \infty$) in which the load of a failed fiber is shared beyond the nearest neighbours but not globally, with the range depending on $\gamma$.

The probability $P_j(t)$ of the breaking of a fiber $j$ in one sweep of the lattice in the time interval $\delta_k$ is given by

$$P_j(t) = \sigma_j^\rho(t)\,\delta_k \qquad (22)$$

$\rho$ is the Weibull index ($2 \le \rho \le 50$) which gives the degree of heterogeneity of the system; as $\rho$ increases the material becomes more homogeneous. $\sigma_j(t)$ is the stress on the fiber $j$ at time $t$. The results of our simulations are presented in Fig. 3.

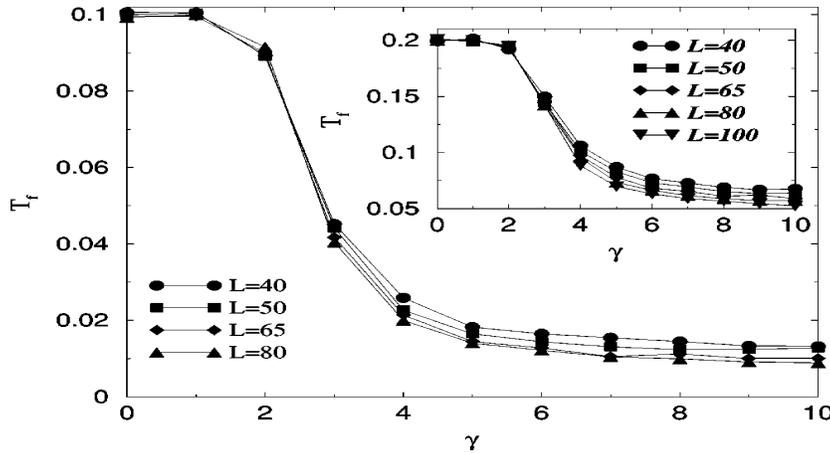

Figure 3: Comparison of the time to failure $T_f$ (dimensionless) obtained for material heterogeneity level $\rho = 10$ (inset $\rho = 5$ ), as the range of interaction $\gamma$ varies. L is the system size.

Our results indicate that the failure time of the material decreases with increasing range of interaction between the composites of the material and this decrease is precipitous when the interaction range is just above next nearest neighbour interaction. Also, our results indicate that the time to failure is delayed if the material is made less heterogeneous; that is, of fewer composites. Again, these results are in excellent agreement with experimental observations [46, 47]

**Conclusions**

This article reviews the modern theory and implementation of computer simulations disordered condensed matter systems and fluids, based on the statistical mechanics approximation method of Markov chain Monte Carlo techniques. We started with the fundamentals without recourse to any system or problem for a generalized treatment and then discussed a number of recent applications including Ising model or spin-glass model of magnetic systems, a solid-on-solid model of nanostructure formation on surfaces driven by ion-bombardment, statistical MC models of material fracture, and a MC model of chirality which is applicable as a matching algorithm. Basic and typical results of our simulations of the different complex systems are presented.

It is hoped that the in-depth treatment of the rudiments of the method, and the excellent agreement of its results in diverse applications would motivate and stimulate its application to the solution of still unsolved statistical physics problems from the perspective of the richer energy landscapes to be encountered in nanometric surface science.

**Acknowledgement**
The author thanks Yamir Moreno, Alexander K. Hartmann, Reiner Kree, Maureen Neal, and Robert Low for discussions.

## References

1. **Landau, David and Binder, Kurt.** *A guide to Monte Carlo simulations in statistical physics.* Cambridge : Cambridge University Press, 2000. And references therein.
2. **Newman, M. E. J. and Barkema, G. T.** *Monte Carlo methods in statistical physics.* Oxford : Oxford University Press, 2002.
3. **Thijssen, J. M.** *Computational Physics.* Cambridge : Cambridge University Press, 1999.
4. **Hartmann, AK and Weigt, M.** *Phase Transitions in Combinatorial Optimization Problems.* Berlin : Wiley-VCH, 2005.
5. **Landau, LD and Lifshitz, EM.** *Statistical Physics: Landau and Lifshitz Course on Theoretical Physics, Vol. 5.* Oxford : Pergamon Press, 1980.
6. **Hartmann, AK and Rieger, H.** *Optimization algorithms in Physics.* Berlin : Wiley-VCH, 2001.
7. **Hartmann, AK and Rieger, H.** *New Optimization Algorithms in Physics.* Berlin : Wiley-VCH, 2004.
8. **Barabasi, A-L and Stanley, HE.** *Fractal Concepts in Surface Growth.* Cambridge : Cambridge University Press, 1995. And references therein.
9. **Ryogo, K, et al.** *Thermodynamics: An advanced course with problems and solutions.*

10. **Press, WH, et al.** *Numerical Recipes: The Art of Scientific Computing.* Third Edition. Cambridge : Cambridge University Press, 2007.

11. **Fortuin, CM and Kasteleyn, PW.** 1972, Physica , Vol. 57, p. 536.

12. **Swendsen, RH and Wang, JS.** 1987, Phys. Rev. Lett., Vol. 58, p. 86.

13. **Wang, J.-S.** 1990, Physica A, Vol. 164, p. 240.

14. **Wolff, U.** 1989, Phys. Rev. Lett., Vol. 62, p. 361.

15. **Baxter, R. J.** *Exactly solved models in statistical mechanics.* London : Academic Press, 1982.

16. *Calculation of partition functions by measuring component distributions.* **Hartmann, AK.** 2005, Phys. Rev. Lett, Vol. 94, p. 050601. And references therein.

17. **Igwe, IE, et al.** *Reports on statistical mechanics theory of the thermodynamic properties of magnetic systems.* Department of Physics, University of Ibadan. Ibadan : s.n., 2010-2011. Postgraduate and undergraduate projects in Theoretical Physics. Contributors: Igwe, IE; Alabi, BO. Atashili, AO; Lawal, LO; Oricha, TO.

18. **Carter, G, Navinsek, B and Whitton, JL.** *Sputtering by Particle Bombardment.* [ed.] R Behrisch. Heidelberg : Springer-Verlag, 1983. p. 231. Vol. II.

19. **Eklund, EA, et al.** 1991, Phy. Rev. Lett., Vol. 67, p. 1759.

20. **Eklund, EA, Snyder, EJ and Williams, RS.** 1993, Surf. Sci., Vol. 285, p. 157.

21. **Bradley, RM and Harper, JME.** 1988, J. Vac. Sci. Technol. A, Vol. 6, p. 2390.

22. *Dynamic Scaling of Ion-Sputtered Surfaces.* **Cuerno, R and Barabasi, A-L.** 1995, Phys. Rev. Lett., Vol. 74, pp. 4746-4749.

23. **Makeev, M, Cuerno, R and Barabasi, A-L.** 2002, Nucl. Instrum. Methods Phys. Res. B, Vol. 197, p. 185.

24. **Hartman, AK, et al.** 2002, Phys. Rev. B, Vol. 65, p. 193403.

25. *Theory of sputtering I. Sputtering yield of amorphous and polycrystalline targets.* **Sigmund, P.** 1969, Phys. Rev., Vol. 184, p. 383.

26. *Propagation of ripples in Monte Carlo models of sputter-induced surface morphology.* **Yewande, EO, Hartmann, AK and Kree, R.** s.l. : Phys. Rev. B, 2005, Vol. 71, pp. 195405(1)-(8).

27. *Morphological regions and oblique-incidence dor formation in a model of surface sputtering.* **Yewande, EO, Kree, R and Hartmann, AK.** s.l. : Phys. Rev. B, 2006, Vol. 73, pp. 115434(1)-(8).

28. *Numerical analysis of quantum dots on off-normal incidence ion sputtered surfaces.* **Yewande, EO, Kree, R and Hartmann, AK.** s.l. : Phys. Rev. B, 2007, Vol. 75, pp. 155325(1)-(8).

29. **Siegert, M and Plischke, M.** 1994, Phys. Rev. E, Vol. 50, p. 917.

30. **Smilauer, P, Wilby, MR and Vvedensky, DD.** 1993, Phys. Rev. B, Vol. 47, p. 4119.

31. **Kelvin, Lord.** *Baltimore Lectures on Molecular Dynamics and the Wave Theory of Light.* London : C. J. Clay and Sons, 1904.

32. **Buda, AB, Auf der Heyde, T and Mislow, K.** 1992, Angew. Chemie. Int. Ed. Engl., Vol. 31, p. 989.

33. **Buda, AB and Mislow, K.** 1992, J. Am. Chem. Soc., Vol. 114, p. 6006.

34. **Solymosi, M, et al.** 2002, J. Chem. Phys., Vol. 116, p. 9875.

35. **Solymosi, M, et al.** 2002, Ferroelectrics, Vol. 277, p. 169.

36. **Neal, M, et al.** 2003, J. Chem. Phys., Vol. 119, p. 3567.

37. **Kamberaj, H, et al.** 2004, Mol. Phys., Vol. 102, p. 431.

38. **Osipov, MA, Pickup, BT and Dunmur, DA.** 1995, Mol. Phys., Vol. 84, p. 6. ibid. 84, 1193 (1995).

39. **Neal, MP.** 2008, Mol. Cryst. Liq. Cryst., Vol. 494, p. 252.

40. *The Hausdorff chirality measure and a proposed Hausdorff structure measure.* **Yewande, EO, Neal, MP and Low, R.** 2009, Mol. Phys., Vol. 107, pp. 281-291.

41. **Herrmann, HJ and Roux, S, [ed.].** *Statistical Models for the Fracture of Disordered Media.* Amsterdam : North Holland, 1990.

42. *Critical Phenomena in Natural Sciences.* **Sornette, D, [ed.].** Berlin : Springer-Verlag, 2000.

43. *Statistical Physics of Fracture and Breakdown in Disordered Systems.* **Chakrabarti, BK and Banguigui, LG, [ed.].** Oxford : Clarendon Press, 1997.

44. *Time evolution of damage under variable ranges of load transfer.* **Yewande, EO, et al.** 2003, Phys. Rev. E, Vol. 68, pp. 026116(1)-(8).

45. **Hidalgo, RC, et al.** 2002, Phys. Rev. E, Vol. 65, p. 046148.

46. *Critical Phenomena in Natural Sciences.* **Sornette, D. [ed.].** Berlin: Springer-Verlag, 2000.

47. *Statistical Physics of Fracture and Breakdown in Disordered Systems.* **Chakrabarti, B. K. and Banguigui, L. G. [ed.].** Oxford: **Clarendon Press**, 1997.